\documentclass[]{interact}

\usepackage{epstopdf}
\usepackage[caption=false]{subfig}

\theoremstyle{plain}

\theoremstyle{definition}

\theoremstyle{remark}

\usepackage{booktabs} 
\usepackage{amsthm}
\usepackage[longnamesfirst,sort]{natbib}
\setcitestyle{aysep={}} 
\bibpunct[, ]{(}{)}{;}{a}{}{,}

\usepackage{graphicx}
\usepackage{float}
\usepackage{adjustbox}
\usepackage{longtable}
\usepackage{rotating}
\usepackage{pdflscape}
\usepackage{hyperref}
\usepackage{xcolor}
\usepackage{graphicx}
\usepackage{multirow}
\usepackage[ruled,vlined]{algorithm2e}

\begin{document}

\articletype{Research articles}

\title{Identifying Risk Variables From Raw ESG Data Using Its Hierarchical Structure}

\author{
\name{Zhi Chen\textsuperscript{a,*}\thanks{* Corresponding author: Zhi Chen. Email: zchen100@stevens.edu}, Zachary Feinstein\textsuperscript{a}, Ionu\c{t}~Florescu\textsuperscript{a,b}}
\affil{\textsuperscript{a}School of Business, Stevens Institute of Technology, 1 Castle Point Terrace, Hoboken, New Jersey, USA;\\ \textsuperscript{b}Hanlon Financial Laboratories, School of Business, Stevens Institute of Technology, 1 Castle Point Terrace, Hoboken, New Jersey, USA}
}

\maketitle

\begin{abstract}
Environmental, Social, and Governance (ESG) data provides non-financial insights into corporations. 
In this study, we aim to identify relevant ESG raw variables to assess financial risk, measured by logarithmic volatility of return.
We propose a framework specifically designed for ESG datasets characterized by a hierarchical data structure and a significantly larger number of variables than observations. 
We show that raw variables selected by the proposed framework are significantly more relevant to financial risk than aggregated ESG scores. Furthermore, these selected risk variables provide additional insights beyond the traditional financial factors. 
We validate the robustness of this framework using out-of-sample data. 
We illustrate our framework using company data from various sectors of the US economy. 
We further identify the specific ESG risk variables relevant to large and small companies within each sector.

\vspace{1em}
\end{abstract}

\section{Introduction}
ESG is a framework designed to provide additional information beyond traditional financial measures collected by corporate financial statements. Many studies tried to use the ESG framework to create better return models \citep{verheyden2016esg, halbritter2015wages}. However, recent literature shows that ESG is in fact more relevant to risk rather than return of a corporation. For example, \cite{engelhardt2021esg,deng2019can,perez2022does, zhou2021esg} find that ESG overall score of a company has a stronger relationship with risk, as measured by volatility of returns. 
\cite{giese2019foundations} explains the ESG-risk relation by introducing a transmission channel between them. They explain that a company with a higher ESG score indicate better risk management, which lead to lower financial risk. In this paper, we explore the ESG-risk relationship.

To assess the companies environment, social and governance exposure, data providers, such as LSEG,\footnote{The London Stock Exchange Group (LSEG), was formerly known as Refinitiv.}, Bloomberg, and MSCI, collect raw ESG related variables from corporation reports and design a hierarchical structure to aggregate this data into a single overall score \citep{halbritter2015wages, shen2023esg, wang2023sparsity}. 
When analyzing the link between ESG and corporate risk, researchers typically rely on these aggregated scores \citep{engelhardt2021esg, zhou2021esg, loof2022corporate}. 
While using aggregated scores is convenient, many studies raise concerns of their limitations \citep{park2024empirical, gibson2021esg}. For example, \cite{erhart2022take} points out that during the averaging process from raw variables, high score in some can offset severe low score in others, ultimately obscuring critical risks in the final score.

To our knowledge, this is the first study to use the complete set of ESG raw data directly into a feature selection method. Prior studies such as \cite{gurvich2022carbon, bolton2023global, fauver2018does} 
rely on domain knowledge to pre-determine one or few raw variables relevant to the response variable analyzed. \cite{bonacorsi2024esg} pre-set rules to filter out approximately 86\% of the available variables. In this study, we systematically use the entire set of ESG raw variables to identify which ESG variables are relevant for financial risk.
To achieve this, we develop a framework specifically for ESG datasets characterized by a hierarchical structure with significantly more variables than observations.
We validate the robustness of our framework through out-of-sample testing. We further validate its superior performance compared to traditional feature selection methods. 

Our results indicate that using selected ESG raw variables has significantly stronger explanatory power than using aggregated scores. More importantly, these selected ESG factors remain relevant even in the presence of traditional financial factors. Within this framework, we can identify risk variables for each industry sector of the US economy. We find that the relevant ESG risk variables for small companies differ significantly from those identified for large corporations even within the same sector. 
This insight suggests that accurately identifying relevant risk variables for a company requires considering both its industry sector and size. 
Many existing studies and regulatory frameworks focus on large companies and identify a single primary risk dimension for each sector \citep{bolton2023global, pastor2022dissecting}. For example, environmental factors are dominant in the Energy sector. In our work, we show that environmental factors indeed dominate for large companies in the Energy sector, while social factors account for a larger proportion of risk for small companies.

The remainder of this work is organized as follows.
Section \ref{sec:whyuselogvol} discusses the measure of risk used in this study. Section \ref{sec:hvsalgorithm} introduces selection processes of risk variables using the Energy sector as an example. In section \ref{sec:HVSvalidation}, we evaluate the framework's robustness and compare it with benchmark methods. Finally, section \ref{sec:CaseStudies} presents case studies: we identify relevant risk variables for each industry sector, as well as for both large- and small-capitalization companies by sector.

\section{Logarithmic volatility as response variable}
\label{sec:whyuselogvol}
We begin our analysis by determining the appropriate response variable for the ESG data. In this work we use the logarithmic of the volatility of returns as a measure for the financial risk of a company. Next, we explain our reasons for choosing this particular measure of risk. 

Many studies construct a regression between the ESG overall score and the return of a corporation. Most studies do not find a significant relationship \citep{engelhardt2021esg, li2021esg, halbritter2015wages}. Due to the lack of correlation, other studies consider alternative return-related response variables such as earnings per share, ROA, or Tobin's Q \citep{deng2019can, fauver2018does, alareeni2020esg}, but these studies also reveal no significant correlation. Similarly, our study does not identify a relation between ESG data and returns even when using raw variables (see Appendix \ref{appendix:applyinghvsonreturns}).

Recent research suggests that ESG data is more relevant to risk-related response variables. \cite{engelhardt2021esg, zhou2021esg, sabbaghi2023esg} find a statistically significant negative correlation between ESG aggregate variables and the volatility of returns. \citep{bonacorsi2024esg} finds a similar relationship with a company's z-score calculated using six financial accounting variables. However, the relationship between ESG variables and volatility may not be linear. \cite{livieri2024pricing} argues that changes in environmental policy or innovation can incur sudden costs that undermine the financial stability of the firm. This may create spikes in volatility, which results in heavily right-skewed residuals. 
\cite{zhang2025impact} shows that ESG is a significant predictor of volatility for low-ESG firms, whereas this predictive power diminishes for high-ESG firms. They argue this occurs because once a firm achieves a high sustainability standard, its risk is increasingly driven by diverse conditions rather than internal governance. This can cause heteroskedasticity, thus violating the assumption of constant variance of residuals for the standard regression.

Consistent with these economic phenomena, our diagnostic checks reveal that the model's residuals are heteroskedastic and the distribution is right-skewed \ref{fig:residual_analysis}. 
To address this problem, we use the Box-Cox transformation to identify an appropriate functional form for the response variable. 
\begin{equation}
y^{(\lambda)} =
\begin{cases}
\frac{y^\lambda - 1}{\lambda}, & \text{if } \lambda \neq 0 \\
\log(y), & \text{if } \lambda = 0
\end{cases}
\end{equation}

The results suggest that a logarithmic transformation ($\lambda = 0$) of volatility satisfies the model assumptions, as illustrated in Table  \ref{tab:Comparison of regular volatility and logarithmic volatility}.
Using logarithmic volatility also leads to an increase in the R-squared value of our framework, indicating improved model fit and explanatory power.

\begin{table}[htbp]
  \centering
  \caption{Comparison of regular volatility and logarithmic volatility}
    \begin{tabular}{rcc}
    \toprule
    Volatility Form & \multicolumn{1}{l}{The p-value of Jarque-} & \multicolumn{1}{l}{$R^2$} \\
    &Bera Test For Residuals&\\
    \midrule
     Volatility & 0.0000 & 0.39 \\
    Logarithmic Volatility & 0.1389 & 0.51 \\
    \bottomrule
    \end{tabular}
  \label{tab:Comparison of regular volatility and logarithmic volatility}
\end{table}

As a result, to quantify risk we calculate the annual logarithmic volatility using daily return observations for each company in the dataset using the following formula: 
\begin{equation}
    \log(s) = \log\left(\sqrt{\frac{1}{N-1} \sum_{t=1}^{N} (r_t - \bar{r})^2}\right)
\end{equation}
where: $N$ is the number of trading days in a year; $r_t$ is the daily return of day t; $\bar{r}$ is the average return over $N$ days.

\section{Methodology used to select relevant risk variables}
\label{sec:hvsalgorithm}
Within this section we introduce our proposed framework that selects a subset of ESG raw variables relevant to the logarithmic volatility of returns. This framework is specifically designed to use the hierarchical structure of the LSEG dataset. We first describe the data structure and then detail the framework. We illustrate our framework using data from companies in the Energy sector.


\subsection{LSEG dataset and its structure}\label{subsec: data structure}
\begin{figure}[htb]
    \makebox[\textwidth][c]{ 
        \begin{minipage}{1\textwidth}  
            \centering
            \caption{Hierarchical Structure of the LSEG ESG Dataset}
            \label{Data Structure of ASSET4}
            \includegraphics[width=\linewidth]{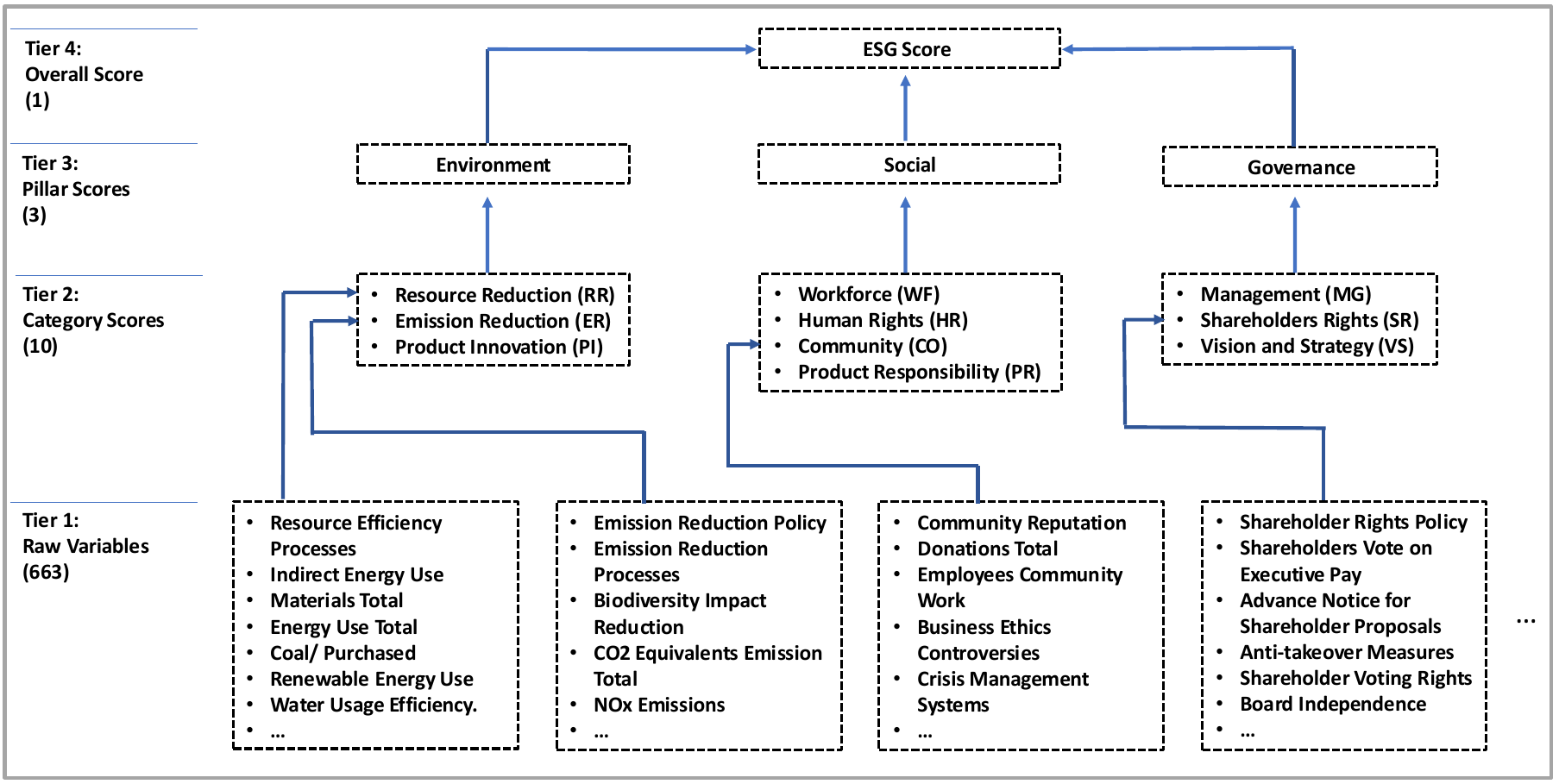}
        \end{minipage}
    }
\end{figure}

We chose to use the LSEG dataset because it provides access to a large number of raw variables. Figure \ref{Data Structure of ASSET4} describes the structure of this dataset. The raw variables are aggregated hierarchically through several tiers. Specifically, the raw scores are aggregated into 10 category scores, which are then combined into 3 pillar scores, and ultimately  into a single overall score, as illustrated by the arrows in Figure \ref{Data Structure of ASSET4}.  

Many of the raw variables in the dataset contain missing values.
We follow \cite{clarkson2008revisiting,cheng2014corporate} by assigning $0$ to missing values for boolean variables. 
We then remove numerical variables that have less than 80\% data available.
Finally, any company-year observation that still contains missing values after these steps is removed. The exact details of data processing steps are provided in Appendix \ref{appendix:dataprocessing}.
Taking the Energy sector as an example, the initial dataset has 617 variables and 695 company-year data points. After the data processing steps, the final dataset for the Energy sector contains 255 variables and 422 observations.

\subsection{Selection process}
Our selection process is motivated by the specific characteristics of ESG raw data. The ESG dataset contains many variables with a limited number of observations. To entice customers, data providers have been expanding the set of ESG raw variables available. For instance, Refinitiv (currently LSEG) advertised the number of raw variables collected as 400 in 2017 and 600 in 2023. 
Increasing the number of features collected may serve as a marketing tool. However, data collected in previous years may not be augmented easily with values for the new variables introduced in later years. Thus, the resulting database is very sparse. This sparsity leads to issues such as multicollinearity among the available variables. Applying feature selection methods using the entire dataset is prone to overfitting. To avoid this issue, we propose a framework that splits the variables into several subsets and selects risk variables from each subset at every stage.


\paragraph*{Stage 1: Selecting raw variables within each category of ESG dataset.}
As introduced in Section \ref{subsec: data structure}, LSEG uses the category level (Tier 2 in Figure \ref{Data Structure of ASSET4}) in order to partition raw variables. Each category is designed to assess exposure to specific corporate activities and strategic objectives \citep{eikon2022environmental}.
For example, \cite{gurvich2022carbon, bolton2023global} select variables within the Emission Reduction category to measure the carbon emission. \cite{fauver2018does} select the variables within the Employment Quality category to proxy good employment. In our work, we first select relevant variables from each category introduced by LSEG. To achieve this, we perform a Stepwise regression (use the AIC criterion for model selection) with raw variables within each category. Note that the specific choice of feature selection method is flexible, as the main purpose of this stage is to select relevant variables from categories that assess specific types of ESG-risk.
This process produces 10 sets of variables for each one of the 10 categories. For the Energy sector example, the total number of selected variables is 55 from an initial 255.

\begin{figure}[h]
    \makebox[\textwidth][c]{
        \begin{minipage}{1\textwidth}  
            \centering
            \caption{This bar chart illustrates the Adjusted R-squared values when using selected raw variables within each category to regress logarithmic volatility of returns. The dashed lines indicate the Adjusted R-squared values obtained from linear regression models using aggregated scores provided by LSEG as regressors (i.e., 10 category scores, 3 pillar scores, single overall score).}
            \label{comparionofcateogystep1}
            \includegraphics[width=\linewidth]{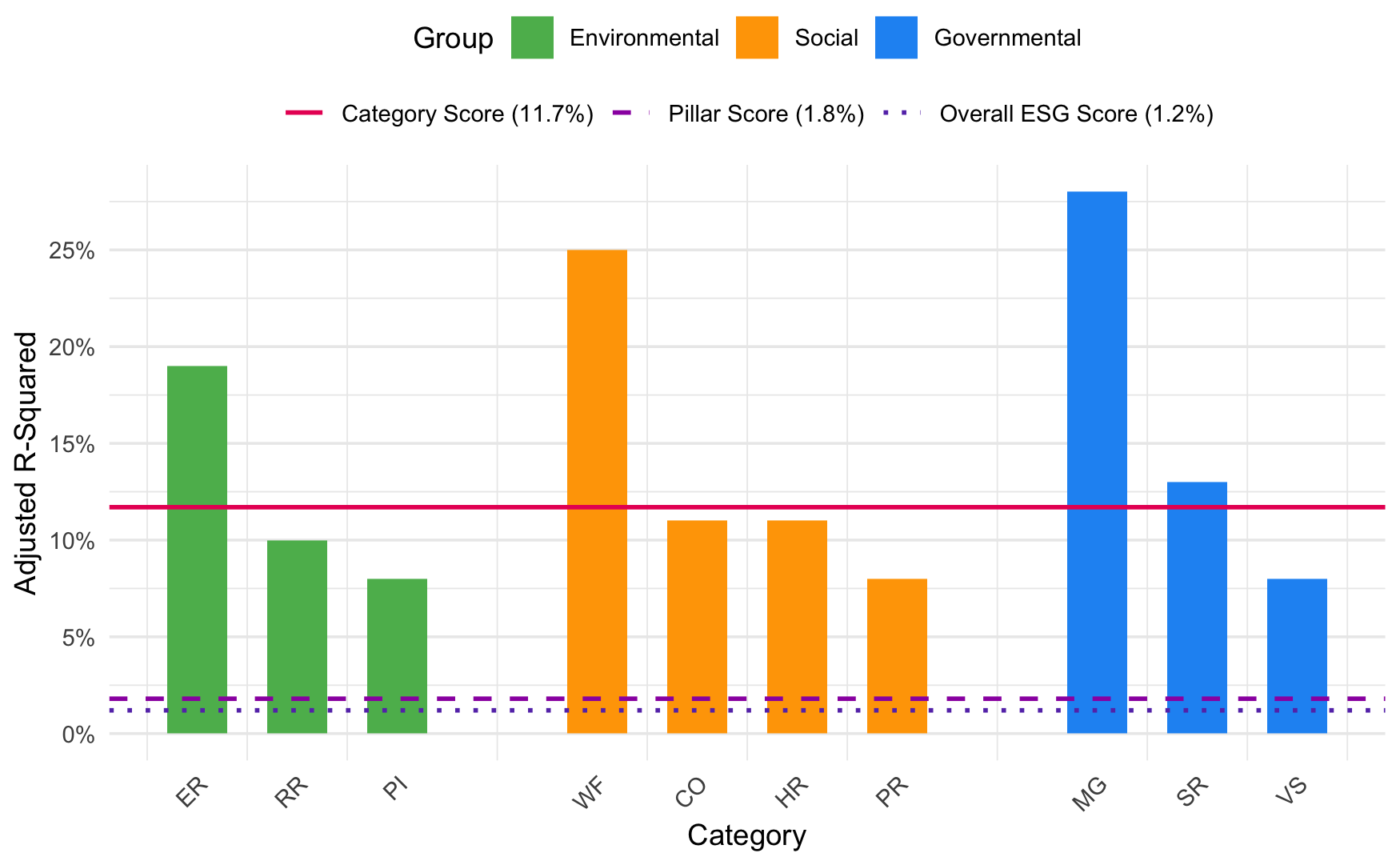}
        \end{minipage}
    }
\end{figure}

The performance of these 10 regressions for the Energy sector is visualized in Figure \ref{comparionofcateogystep1} on page \pageref{comparionofcateogystep1}. The bars represent the Adjusted R-squared values for each regression. We use Adjusted R-squared values for visualization. For comparison, the plot also includes three horizontal lines representing the Adjusted R-squared values from benchmark models, each using one of LSEG's aggregated score (overall, pillar, or category) as predictors.

Looking at the three lines in Figure \ref{comparionofcateogystep1}, we see an indication already that using more detailed ESG scores leads to better model performance even when considering adjusted R-squared which penalizes for having more variables in the model. 
A more important note is to see that models built using selected raw variables from certain specific ESG categories\footnote{Emission Reduction from the E pillar, Workforce from the S pillar, Management and Shareholders Rights from the G pillar} outperform the model that uses all 10 pre-aggregated category scores. This is particularly striking given that these 10 category scores are aggregated from the entire set of raw variables. 
This finding suggests that the aggregation process itself obscures or dilutes the impact of the most critical risk-relevant variables.

\paragraph*{Stage 2: Selecting variables from all significant variables identified in Stage 1 across categories.} In this stage we further refine the variables selected in previous stage. We aggregate all variables selected by the 10 regressions in the previous stage. However, simply adding them together could introduce multicollinearity. To mitigate this, we perform a second Stepwise regression with these selected variables.

\begin{figure}[h]
    \centering
    \includegraphics[width=1\linewidth]{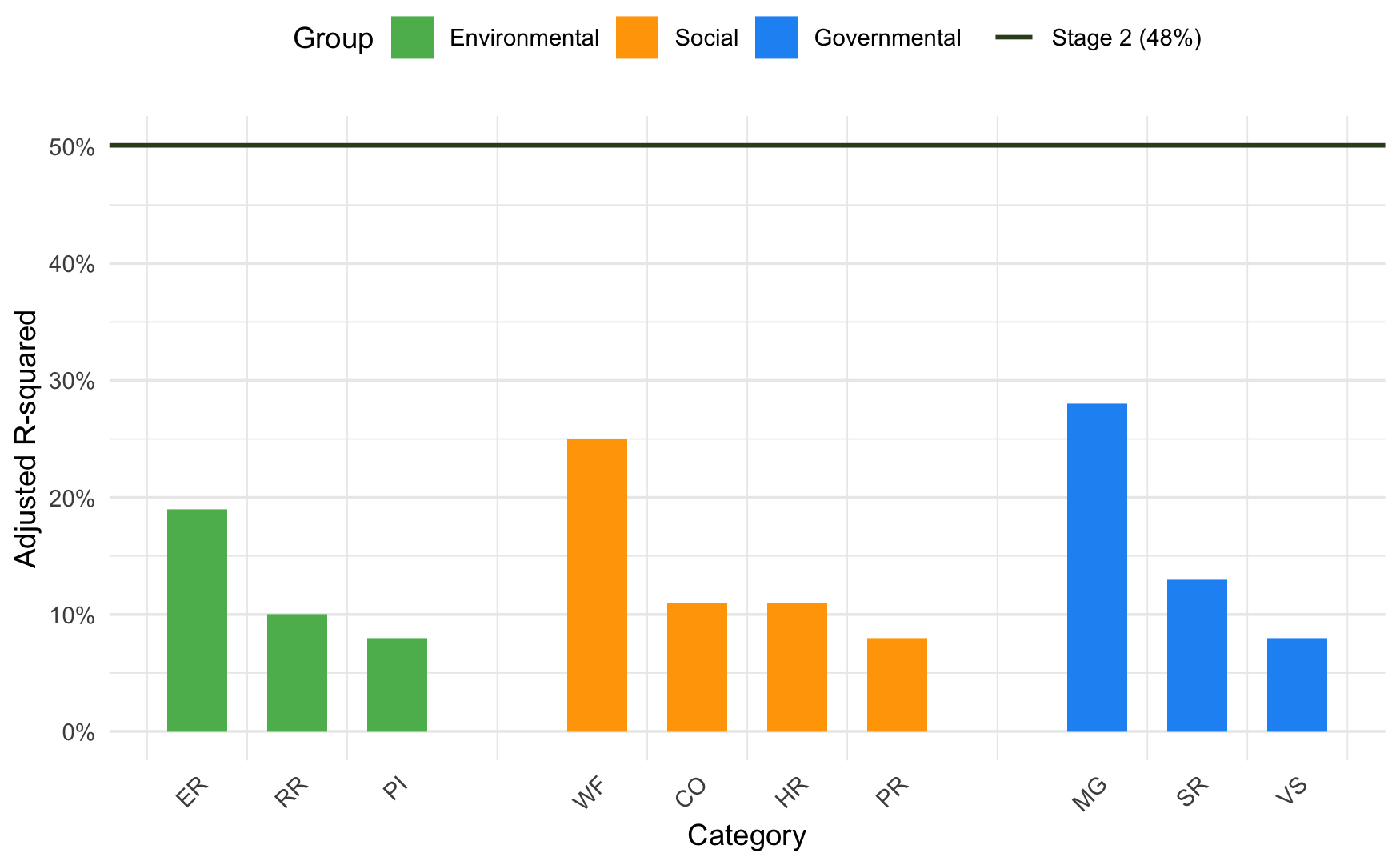}
    \caption{Adjusted R-squared values of stage 1  vs. stage 2 for the Energy Sector}
    \label{fig:resultstep1vsstep2}
\end{figure}

The Adjusted R-squared value from this consolidated model is significantly higher than the values achieved by any of the models using selected variables from a single category (Figure \ref{fig:resultstep1vsstep2}).
This indicates that variables from different categories provide diverse and complementary contributions to the ESG risk model. To our knowledge, this Adjusted R-squared value is also larger than those reported in other studies \citep{engelhardt2021esg,zhou2021esg, loof2022corporate}. 
In addition to increasing performance, this stage further selects variables most relevant to logarithmic volatility of return. For example, in the Energy sector the number of variables is reduced from 51 to 35 and at the same time the Adjusted R-squared value increases to around 48\%. 

\paragraph*{Stage 3: Enhancing the model's robustness} The objective of Stage 3 is to enhance the model's robustness. Given that ESG datasets typically contain a large number of variables but relatively few observations, there is a high risk of overfitting. This step is focused on improving the model's out-of-sample performance. We perform a Ridge regression with the selected variables identified in Stage 2.
We use Ridge regression \citep{Hoerl01021970} as it has a penalty term that reduces the risk of overfitting the dataset. We use a 5-fold cross validation technique to determine the regularization parameter.
Additionally, we standardize all variables, as this will enable us to compare the effect of each variable on the response.

\begin{table}[htbp]
  \centering
  \caption{Comparison of Model Fit Metrics after Stage 2 (Stepwise) and Stage 3 (Ridge) for the Energy Sector} \label{tab:Results of HVS Step 2 and HVS Step 3}
  
    \begin{tabular}{lrrrr}
    \toprule
    Stage  & \multicolumn{1}{l}{No.Selected Variables} & \multicolumn{1}{l}{\%dev} & \multicolumn{1}{l}{AIC} & \multicolumn{1}{l}{BIC} \\
    \midrule
    Stage 2 & 35    & 0.52  & 344.65 & 491.94 \\
    Stage 3 & 35    & 0.51   & 351.19 & 496.39 \\
    \bottomrule
    \end{tabular}
\end{table}

This final step retains the same set of variables selected in Stage 2, but with weights modified through Ridge regression for enhanced robustness. We provide the definitions of all variables identified as important for the Energy sector in \ref{appendix:definitionofriskvariables}. 
Table \ref{tab:Results of HVS Step 2 and HVS Step 3} shows that for the Energy sector, the \%dev\footnote{The Ridge regression cannot calculate an $R^2$ value. Instead, it calculates the Percent Deviance Explained \%dev \citep{Hoerl01021970} to refer to the proportion of deviance explained by the Ridge regression. In the case of multivariate regression \%dev is exactly equal to the $R^2$ value.} value obtained in Stage 3 is nearly identical to the value obtained in Stage 2. In fact all statistics are very similar. However, this final step is crucial as it creates a more robust methodology as we will show in the next section. 

\begin{figure}[htp]
\caption{This pie chart illustrates the sum of absolute value of coefficients for selected raw variables within each category resulted from Stage 3 for the Energy sector companies.}
\label{Significnt_Variables_Distribution}
\centering
\includegraphics[width=0.85\textwidth]{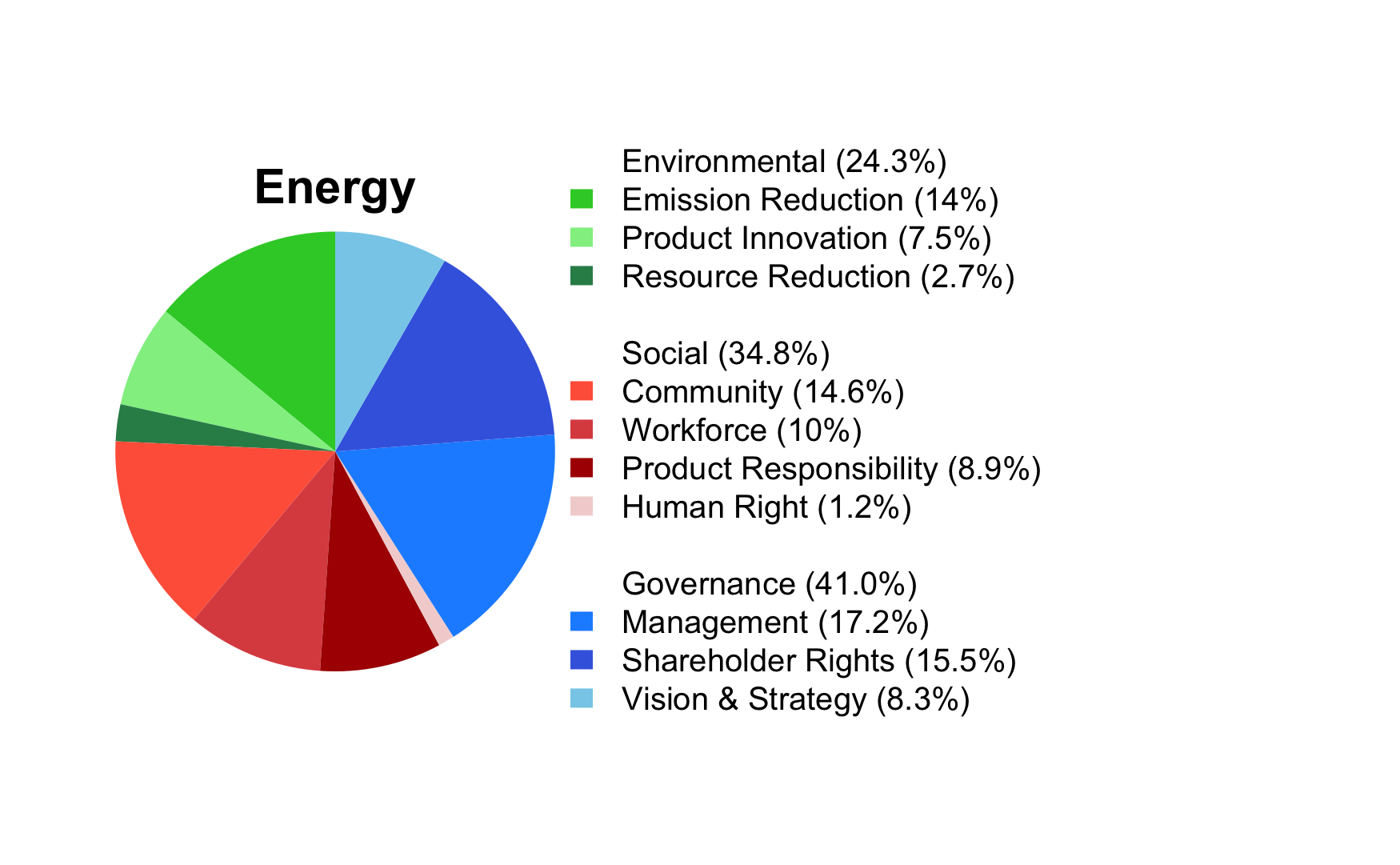}
\end{figure}

More importantly, we are able to assess the relative importance of selected variables for logarithmic volatility of returns, thus highlighting which ESG risks are more relevant. 
\phantomsection
\label{para:calculate_pie_chart}
In Figure \ref{Significnt_Variables_Distribution} we visualize the importance of the 10 ESG categories of the Energy sector. 
Let the final model from Stage 3 be:
\begin{equation}
    \text{Logarithmic Volatility} = \beta_0 + \sum_{i=1}^{k} \beta_i x_i + \epsilon
\end{equation}
where $k$ is the total number of selected variables, and the predictor variables ($x_i$) have been standardized. The coefficient $\beta_i$ quantifies the impact of variable $i$ on logarithmic volatility and since the variables are standardized they have comparable magnitudes. Each of the $k$ selected variables belongs to one of the 10 specific ESG categories ($C_1,C_2,\dots,C_{10}$). To determine the importance of each category we sum the absolute values of the $\beta_i$’s of variables from that category. Specifically, if $I_j$ denotes the set of indices for variables in category $C_j$ we have:
\begin{equation}
    \text{Score}_{C_j} = \sum_{i \in I_j} \lvert\beta_i\rvert
\end{equation}
For example, if variables $x_1$, $x_5$, and $x_{12}$ belong to the Community category, its total influence score is calculated as $\lvert\beta_1\rvert + \lvert\beta_5\rvert + \lvert\beta_{12}\rvert$. This is done for all 10 categories.
Finally, the importance of each category $C_j$ is calculated as its relative score expressed as a percentage:
\begin{equation}
    \% \text{ Importance}_{C_j} = \frac{\text{Score}_{C_j}}{\sum_{n = 1}^{10} \text{Score}_{C_n}} \times 100\% = \frac{\sum_{i \in I_j} \lvert\beta_i\rvert}{\sum_{i=1}^{k} \lvert\beta_i\rvert} \times 100\%
\end{equation}


Figure \ref{Significnt_Variables_Distribution} displays a pie chart of importance weights for each category in the Energy sector.
The colors are chosen such that all environment (E) categories are shades of green, social (S) categories are shades of red, and governance (G) are shades of blue. Each category is assigned a fixed color.

Intuitively, one may expect environmental variables to dominate the other categories for companies in the Energy sector. However, we can clearly see that variables related to governance categories account for the most significant part of ESG-related risk. This seemingly counterintuitive result will be addressed in Section \ref{ssec:sectorbysize} when we study each sector factored by the size of the company. 

We are going to validate this framework in the next section. This framework is hereafter referred to as Hierarchical Variable Selection (HVS).

\section{HVS Algorithm validation}\label{sec:HVSvalidation}
In this section we illustrate the robustness of the HVS framework. As mentioned in the discussion of Stage 3, the nature of ESG data (few observations, large number of variables) makes overfitting easy. 
Additionally, since HVS selects ESG variables related to changes in logarithmic volatility, we also compare it with other factor selection methods. 
As in Section~\ref{sec:hvsalgorithm}, we illustrate our results using the Energy sector. The analysis, conducted across all sectors, yields consistent findings. Comprehensive results for all other sectors are provided in Appendix \ref{appendix:validationforothersectors}. 

\subsection{In-sample and out-of-sample analysis}
In order to assess the robustness of the methodology we introduce two out-of-sample validation designs. The first design evaluates temporal robustness by testing whether the framework can learn a model from the historical data and use it to predict future logarithmic volatility of equity returns. In this design, we train the HVS model using data from a five-year historical window to predict the logarithmic volatility of corporate returns for the next year. The five-year window is rolled forward annually through the dataset. In \cite{golbayani2020application} the authors show that this temporal data split is much more relevant for financial data than the traditional cross-validation method.  

The second validation design assesses cross-sectional robustness using leave-one-out-company validation within each sector. We iteratively train the HVS model on data from all companies within the Energy sector except one, and then use the trained model to predict the logarithmic volatility for the held-out company. This process is repeated for each company in the sector. For both the temporal and cross-sectional validation scenarios, model prediction accuracy is evaluated using Mean Squared Error (MSE).

To illustrate the results, we compare the performance of the full HVS model with the intermediate model that terminates after Stage 2 (i.e., without the final Ridge regression stage). 
We also include a benchmark model that does not include any ESG variables, where the forecast is simply the average value of the in-sample logarithmic volatility. Specifically, when forecasting the next year logarithmic volatility we use the past five-year average logarithmic volatility for the respective company. Table \ref{tab:Comparison between HVS step 2 and HVS step 3} presents the MSE values as well as statistical tests for both designs.

\begin{table}[htbp]
  \centering
  \caption{Performance and Statistical Comparison of HVS Methods for the Energy Sector}
    \begin{tabular}{rlrr}
    \toprule
    \multicolumn{4}{c}{\textbf{Panel A: Comparison of MSE}} \\
    \midrule
    \multicolumn{1}{l}{Design} & Model & \multicolumn{1}{l}{IS MSE} & \multicolumn{1}{l}{OOS MSE} \\
    \midrule
    \multicolumn{1}{l}{Cross-sectional} & Mean of Response & 0.1871 & 0.2438 \\
          & HVS Stage 2 & 0.1134 & 0.2122 \\
          & HVS Stage 3 & 0.1290 & 0.1839 \\
    \midrule
    \multicolumn{1}{l}{Temporal} & Mean of Response & 0.1871 & 0.2438 \\
          & HVS Stage 2 & 0.0931 & 0.2413 \\
          & HVS Stage 3 & 0.1130 & 0.2132 \\
    \midrule
    \midrule
    \multicolumn{4}{c}{\textbf{Panel B: Statistical Tests (P-Values)}} \\
    \midrule
    \multicolumn{1}{l}{Design} & Comparison of MSE & \multicolumn{1}{l}{IS P-Value} & \multicolumn{1}{l}{OOS P-Value} \\
    \midrule
    \multicolumn{1}{l}{Cross-sectional} & HVS Stage 3 $<$ Mean of Response & 0.0000 & 0.0000 \\
          & HVS Stage 3 $<$ HVS Stage 2 & 1.0000 & 0.0060 \\
          & HVS Stage 2 $<$ Mean of Response & 0.0000 & 0.0933 \\
    \midrule
    \multicolumn{1}{l}{Temporal} & HVS Stage 3 $<$ Mean of Response & 0.0000 & 0.0094 \\
          & HVS Stage 3 $<$ HVS Stage 2 & 1.0000 & 0.0332 \\
          & HVS Stage 2 $<$ Mean of Response & 0.0000 & 0.4594 \\
    \bottomrule
    \end{tabular}
  \label{tab:Comparison between HVS step 2 and HVS step 3}
\end{table}

We can see from Panel A in Table \ref{tab:Comparison between HVS step 2 and HVS step 3} that the full HVS model achieves lower out-of-sample MSE values than the intermediate model that terminates after Stage 2.
To test whether the improvement is statistically significant we perform matched pairs tests using every combination of models. The p-values of these tests are presented in Panel B of Table \ref{tab:Comparison between HVS step 2 and HVS step 3}. These results provide statistical evidence that applying the Ridge regression in Step 3 of the HVS algorithm is important for the robustness of the algorithm when assessing out-of-sample data.

\subsection{HVS versus traditional variable selection methods}
To assess whether the multi-stage variable selection procedure of the HVS framework is justified, we benchmark its performance against four classical methods. 
\begin{enumerate}
    \item We use a Principal Component Analysis (PCA) and we select components that explain $80$\% of the total variance (PCA$_{1}$). 
    \item We use a PCA which is constrained to use the same number of components as the number of variables selected by HVS (PCA$_2$). 
    \item The ``Stepwise'' benchmark uses Stepwise selection but instead of using the Hierarchical approach in HVS it selects all relevant variables in one step.
    \item The ``Lasso'' benchmark uses all raw variables and imposes a LASSO type constraint.    
\end{enumerate}

\begin{table}[htbp]
  \centering
  \caption{Comparison of HVS Performance with Benchmark Methods.}
    \begin{tabular}{lrrrr}
    \toprule
    \textbf{Benchmark} & \multicolumn{1}{l}{\textbf{No.Selected Variables}} & \multicolumn{1}{l}{\textbf{\%dev}} & \multicolumn{1}{l}{\textbf{AIC}} & \multicolumn{1}{l}{\textbf{BIC}} \\
    \midrule
    PCA$_{1}$   & 81    & 0.40  & 450.66 & 790.24 \\
    PCA$_{2}$   & 31    & 0.30  & 471.84 & 610.95 \\
    Stepwise & 60    & 0.56  & 288.11 & 537.68 \\
    Lasso & 72    & 0.46  & 388.95 & 687.61 \\
    \midrule
    HVS   & 35    & 0.51  & 347.76 & 480.68 \\
    \bottomrule
    \end{tabular}
  \label{tab:Models Comparison}
\end{table}

Table \ref{tab:Models Comparison} presents model statistics for each of the selection methods. HVS selects a much smaller number of relevant variables with a comparably high \%dev value. Based on \%dev and AIC, the Stepwise model has a better performance using in-sample data at the cost of doubling the number of selected variables; this may indicate multicollinearity and overfitting.

This is evident in Table \ref{tab:In-Sample and Out-of-Sample MSE for HVS and Stepwise Regression} where we present the MSE values for HVS and the Stepwise method for out-of sample data. While Stepwise regression shows a better fit to the in-sample data, its MSE values increase significantly for out-of-sample data. 
These results indicate that the HVS framework reliably identifies relevant variables while keeping strong robustness on out-of-sample data.

\begin{table}[htbp]
  \centering
  \caption{In-Sample and Out-of-Sample MSE for HVS and Stepwise Regression.}
    \begin{tabular}{rlrr}
    \toprule
    \multicolumn{1}{l}{Design} & Model & \multicolumn{1}{l}{IS MSE} & \multicolumn{1}{l}{OOS MSE} \\
    \midrule
    \multicolumn{1}{l}{Cross-sectional} 
          & Stepwise & 0.0802 & 0.3594 \\
          & HVS   & 0.1296 & 0.1833 \\
    \midrule
    \multicolumn{1}{l}{Temporal} 
          & Stepwise & 0.0571 & 1.5900 \\
          & HVS   & 0.1235 & 0.2124 \\
    \bottomrule
    \end{tabular}
  \label{tab:In-Sample and Out-of-Sample MSE for HVS and Stepwise Regression}
\end{table}

\section{Case study\label{sec:CaseStudies}}
This section illustrates the applicability of the proposed methodology using two case studies. 
First, we compare the model performance using selected raw variables with those using aggregated ESG scores within each industry sector.
We also explore whether these identified risk variables provide additional information beyond traditional financial risk factors.
Second, we examine how relevant ESG risk variables differ by company size within each sector. Most existing literature analyzes the impact of ESG factors for large capitalization companies, for example those corporations included in S\&P 500, China A-shares companies, and the STOXX Europe 600 \citep{alareeni2020esg, deng2019can}. We aim to determine if the same analysis may be applied to small capitalization companies or the relevant ESG factors are different for smaller companies.



\subsection{Sector analysis}
Literature provides clues that using more detailed ESG scores can improve the explanatory power to the response variable analyzed. These findings are based on different datasets and various stock universes \citep{engelhardt2021esg, alsaadi2017corporate, bonacorsi2024esg}. To our knowledge, this is the first study to systematically compare ESG scores across all tiers of the hierarchical data structure. 
Our results reveal that models using selected raw variables consistently and significantly outperform those using aggregated scores within each industry sector.

\begin{figure}[htp]
\caption{A comparison of the \%dev values for HVS with corresponding \%dev values using aggregated LSEG scores.} 
\label{sectoroverallhvscomparison}
\centering
\includegraphics[width=1.0\textwidth]{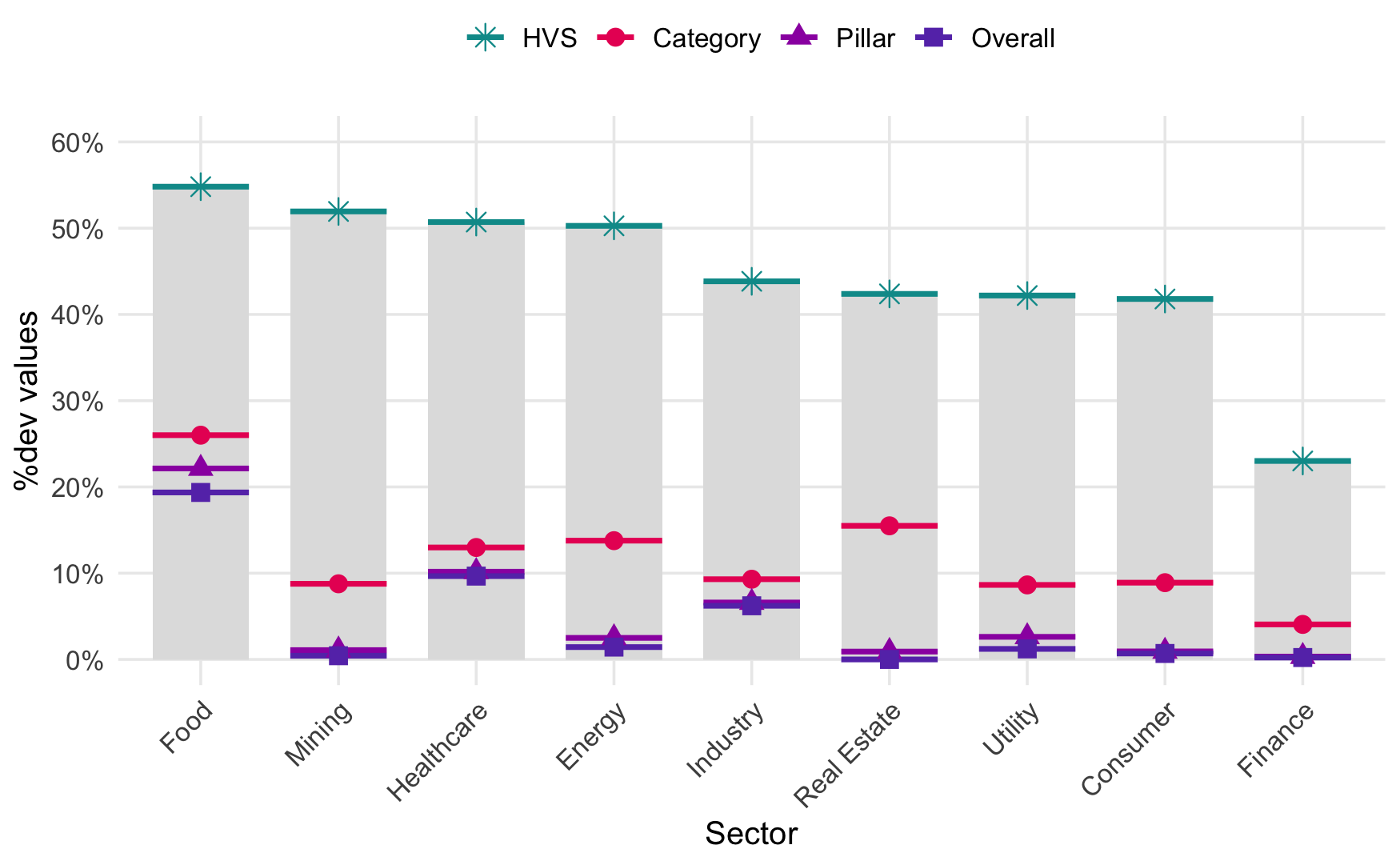}
\end{figure}

In Figure \ref{sectoroverallhvscomparison} we present the \%dev values for HVS (dark cyan) and \%dev values obtained from Ridge regressions of logarithmic volatility with LSEG's aggregated scores. This plot shows that the HVS performance which is based on selected raw variables is universally better than using any of the aggregated scores. 
This figure also illustrates heterogeneity across sectors. Different industries operate under distinct business models, and face different regulations. Thus their exposure and sensitivity to ESG risks vary considerably.
Among the analyzed industries, the Financial sector exhibits the lowest explanatory power. It is important to note that these sector-level results contain both large- and small-capitalization companies.
We revisit these findings in Section \ref{ssec:sectorbysize} where we further explore ESG-risk differences between companies based on size. 


To further show the necessity and robustness of using raw variables and HVS framework, Table \ref{tab:aggregated score coefficients} presents regression coefficients obtained when using pillar and category scores. We can see that statistically significant variables are very sparse, even within the in-sample dataset. We can imagine their predictive power would be even weaker in an out-of-sample testing. In contrast, using HVS allows us to reliably obtain results and draw conclusion. 

\begin{table}[htbp]
  \centering
  \caption{This table reports the coefficients obtained by regressing aggregated ESG scores (three Pillar variables E,S,G) on the logarithmic volatility of returns for companies within each industry sector. In addition, we also report coefficients from models using aggregated ESG scores from Category tier.
  Asterisks (*) denote statistically significant coefficients.}
  \setlength{\tabcolsep}{4pt} 
  \resizebox{\textwidth}{!}{ 
    \begin{tabular}{lrrrrrrrrrrrrr}
    \toprule
          & \multicolumn{3}{c}{Pillar} & \multicolumn{10}{c}{Category} \\
          \cmidrule(lr){2-4} \cmidrule(lr){5-14}
          & \multicolumn{1}{l}{E} & \multicolumn{1}{l}{S} & \multicolumn{1}{l}{G} & \multicolumn{1}{l}{RR} & \multicolumn{1}{l}{ER} & \multicolumn{1}{l}{PI} & \multicolumn{1}{l}{CO} & \multicolumn{1}{l}{HR} & \multicolumn{1}{l}{WF} & \multicolumn{1}{l}{PR} & \multicolumn{1}{l}{MG} & \multicolumn{1}{l}{SR} & \multicolumn{1}{l}{VS} \\
    \midrule
    Food  & -0.09 & -0.33* & 0.01  & -0.03 & 0.11  & 0.05  & -0.29* & 0.14  & -0.29 & 0.12  & 0.01  & 0.01  & -0.28 \\
    Mining & -0.68 & 1.40* & -0.13 & -1.08* & 0.49  & 0.25  & 0.19  & 0.34  & 0.73* & 0.11  & -0.09 & -0.31 & 0.03 \\
    Healthcare & 0.10  & -0.24* & -0.27* & -0.19 & 0.28* & -0.04 & -0.09 & -0.04 & -0.16 & 0.06  & -0.19* & 0.06  & -0.05 \\
    Energy & -0.22 & 0.27  & 0.20  & -0.08 & -0.51* & 0.11  & -0.03 & 0.17  & 0.39  & -0.04 & 0.07  & 0.25  & 0.07 \\
    Industry & -0.03 & 0.00  & -0.24* & -0.12 & 0.06  & 0.02  & 0.02  & -0.01 & -0.04 & 0.00  & -0.17* & 0.00  & 0.03 \\
    Real Estate & 0.05  & 0.47  & -0.36 & -0.24 & -0.16 & 0.07  & 0.06  & 0.44  & -1.04 & 0.37  & -0.31 & -0.07 & 0.22 \\
    Utility & -0.25 & 0.10  & 0.28* & -0.17 & -0.09 & 0.06  & 0.04  & 0.21* & -0.07 & 0.04  & 0.30* & 0.01  & -0.16 \\
    Consumer & -0.33 & 0.40* & -0.06 & 0.26  & -0.19 & -0.29* & 0.63* & -0.19 & -0.29 & 0.24  & 0.11  & -0.04 & 0.00 \\
    Finance & 0.18  & -0.37 & -0.09 & -0.35 & 0.06  & 0.18  & -0.01 & -0.07 & -0.07 & 0.02  & -0.09 & 0.32* & 0.21 \\
    \bottomrule
    \end{tabular}
  }
  \label{tab:aggregated score coefficients}
\end{table}


\paragraph*{The selected raw ESG variables bring new information to traditional financial risk factors.}
Traditional financial risk factors measure companies' profitability, solvency, liquidity and other abilities. Specifically, \cite{wei2006did} find significant negative relationship between a firm's Return-on-Equity (ROE) and the volatility of its stock returns. Other studies also identify risk factors related to leverage ratio, turnover, dividend payout, cashflow, etc \citep{al2021firm, pae2018idiosyncratic}. As a result, we consider the following financial risk factors: leverage ratio, turnover, size, cash ratio, price-to-earnings ratio, book-to-market ratio, dividend payout, return-on-equity(ROE), return-on-asset(ROA), free cash flow, cash flow to investment, net debt, total current liabilities, earnings-per-share growth, operation profit.

A natural question to ask is whether any of the ESG information is already contained in these traditional financial risk factors. To answer this question we compare three regression models for each sector, all of which use logarithmic volatility as the response variable. These consist of (1) a baseline financial risk factor model, (2) a model containing only the HVS-selected ESG variables, and (3) an augmented model that integrates both the financial and ESG factors.

Table \ref{tab:vsfamafrenchthreefactorvol} presents the model fit statistics for this analysis. Augmenting the baseline model with HVS-selected variables substantially improves its performance across every sector, as shown by a significant increase in the \%dev and a corresponding decrease in the BIC. This finding indicates that the ESG factors identified by HVS provide additional information for explaining the volatility of returns.

\begin{table}[htbp]
  \centering
  \caption{Assessing the incremental explanatory power of HVS-selected ESG variables beyond the financial risk factors, by Sector. The values represent the \%dev from regressions on logarithmic volatility of returns.}
  \resizebox{\textwidth}{!}{%
    \begin{tabular}{l*{6}{c}}
    \toprule
          & \multicolumn{2}{c}{Financial risk factors} & \multicolumn{2}{c}{Selected ESG Variables} & \multicolumn{2}{c}{Combined Model} \\
    \cmidrule(lr){2-3} \cmidrule(lr){4-5} \cmidrule(lr){6-7}
    Sector & \%dev & BIC & \%dev & BIC & \%dev & BIC \\
    \midrule
    Food  & 0.25  & 349.41 & 0.55  & 232.90 & 0.60  & 225.55 \\
    Mining & 0.24  & 195.60 & 0.53  & 155.10 & 0.61  & 122.69 \\
    Healthcare & 0.23  & 746.09 & 0.51  & 625.76 & 0.56  & 541.20 \\
    Energy & 0.32  & 492.16 & 0.51  & 480.68 & 0.62  & 368.21 \\
    Industry & 0.28  & 664.44 & 0.44  & 647.10 & 0.54  & 517.75 \\
    Real Estate & 0.24  & 271.68 & 0.42  & 242.98 & 0.55  & 240.93 \\
    Utility & 0.28  & 481.92 & 0.42  & 433.93 & 0.52  & 336.96 \\
    Consumer & 0.26  & 434.35 & 0.42  & 418.82 & 0.51  & 385.53 \\
    Finance & 0.16  & 905.08 & 0.23  & 874.09 & 0.33  & 795.61 \\
    \bottomrule
    \end{tabular}
  }
  \label{tab:vsfamafrenchthreefactorvol}
\end{table}

\subsection{Sector by size analysis \label{ssec:sectorbysize}}
As far as the authors are aware, most prior ESG studies \citep{alareeni2020esg, jain2019can,shen2023esg} focus on large capitalization companies defined as total market value in excess of \$10 billion. At the time of writing this manuscript, this includes all S\&P 500 companies. In this section we focus on the difference in factors for large and small capitalization companies within each sector. We consider companies in the S\&P 500 index to represent large cap companies and companies in the Russell 2000 index to represent small cap companies. We avoid mid cap companies\footnote{i.e.,  companies whose market cap is between these large-cap and small-cap groups} to emphasize differences in ESG variables based on market cap. Since we already know that sectors behave differently, we split companies within each sector into two subsets based on their market capitalization. 


We begin by examining the relevance of ESG factors to corporate volatility based on firm size.
Table \ref{tab:Comparisonr2forlargeandsmallbysector} shows that in every sector analyzed, the fit (quantified by the  \%dev value) is substantially better for large-cap group than the small-cap group. This suggests that the ESG risk factors are more relevant for the volatility of large-cap firms than to that of small-cap firms. 
The Finance sector provides the starkest illustration of the difference between large- and small-cap firms. Further examination reveals that the number of small companies is approximately four times greater than the number of large companies within this sector. This may explain the low \%dev values obtained for the Finance sector in the previous sector analysis (Figure \ref{sectoroverallhvscomparison}). The companies analyzed in the Finance sector tend to be small, thus ESG selected factors are less relevant for their logarithimic volatility.


\begin{table}[htbp]
  \centering
  \caption{Comparison of \%dev Values of HVS algorithm for Large- and Small-Cap Companies by Sector}
  \label{tab:Comparisonr2forlargeandsmallbysector}
  \resizebox{\textwidth}{!}{%
    \begin{tabular}{lccccccccc} 
    \toprule
              & Food  & Mining & Healthcare & Energy & Industry & Real Estate & Utility & Consumer & Finance \\
    \midrule
    Large-cap & 0.50  & 0.55  & 0.37  & 0.48  & 0.42  & 0.47  & 0.36  & 0.49  & 0.52 \\
    Small-cap & 0.38  & 0.33  & 0.32  & 0.39  & 0.39  & 0.37  & 0.24  & 0.29  & 0.06 \\
    \bottomrule
    \end{tabular}
  }
\end{table}

\begin{figure}
    \centering
    \includegraphics[width=1\linewidth]{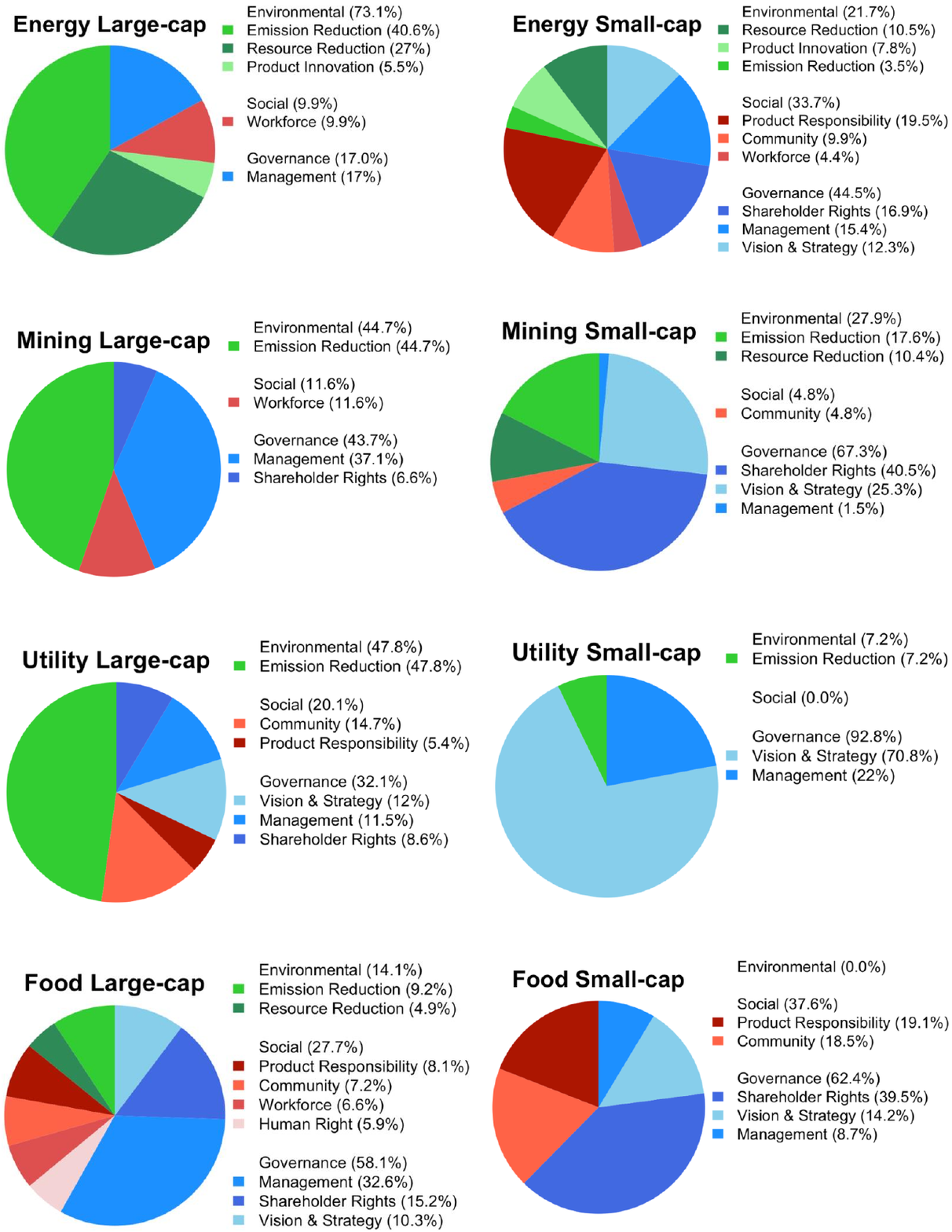}
    \caption{ESG variables within each sector for large and small cap companies.}
    \label{fig:largesmallbysector}
\end{figure}

We want to reiterate that our method identifies specific raw factors and also identifies their relative importance to each other. 
To provide a comprehensive overview of the important factors, we aggregate the magnitude of the estimated coefficients for the selected variables within their respective categories.
Figures \ref{fig:largesmallbysector} present a sector-by-sector comparison of the relative importance of ESG categories for large- and small-cap companies. The importance of variables differs significantly between small and large companies, even within the same sector. For instance, in the Energy sector, environmental variables are the overwhelmingly dominant risk drivers for large-cap companies, with factors such as Emission Reduction and Resource Reduction being particularly significant. Most studies focus on analyzing the impact of ESG factors for large-cap companies, and they found the Environmental factors is significant \citep{konar2001does, matsumura2014firm, bolton2023global, pastor2022dissecting}.

For small Energy companies, in contrast, social and governance factors such as Community and Shareholder Rights take precedence. We observe similar patterns in other resource-intensive sectors, including Mining, Utility, and Food. Environmental factors carry significantly more weight for large-capitalization companies, whereas the dominant relevant risk factor of small-capitalization firms is Governance. In Appendix \ref{fig:pattern2}, we provide figures of the remaining sectors for reader's reference.

The finding that the Environment factors show significant relevance to ESG risk across large-cap companies aligns with the frameworks of legitimacy theory. Legitimacy theory \citep{patten1991exposure, guthrie2007legitimacy} states that firms operate under an implied social legitimacy, defined as the congruence between their actions and societal norms. This explains the importance of the Emission Reduction category for emission-intensive sectors. Legitimacy theory also states that larger companies are subject to more intense scrutiny and social pressure from regulators, investors, and the public. Thus, they would generally face greater exposure to regulatory penalties, environmental litigation, and reputational risk associated with environmental performance \citep{benvenuto2023systematic}. This explains the different importance placing on related variables.

For small-cap companies, Governance is the primary factor. Stakeholder theory \citep{gray1995corporate} suggests that a corporation's continued existence depends on managing relationships with groups including communities, suppliers, and shareholders to mitigate risk. 
Additionally, small-cap companies tend to operate more locally. They often seek to gain a competitive advantage through novel products or quickly take market positioning. Thus, robust strategic planning and strong internal governance structures become important to their stability.

These findings highlight a crucial insight for ESG risk modeling. When identifying relevant ESG risk variables for a specific company, instead only relying on its industry classification, we should further consider its size. When an entire sector is analyzed in aggregate, the resulting dataset becomes more heterogeneous. Consequently, powerful but group-specific variables may be less dominant overall.

\section{Conclusions}
In this study we provide evidence that ESG variables are related to financial risk, as measured by logarithmic volatility. We developed the Hierarchical Variable Selection (HVS) framework specifically to understand ESG datasets. Using raw variables achieves higher explanatory power than using aggregated ESG scores. Compared with traditional selection models, our framework achieves superior performance in modeling risk while identifying a more parsimonious set of variables.

Applying HVS across multiple sectors, we find that different sectors have different relevance to ESG risk. Furthermore, our results reveal that the relevant ESG risk factors differ between large- and small-cap companies, reflecting variations in public expectations, regulatory exposure, and strategic positioning. 

Our results also illustrate the applicability of the proposed framework. Across multiple sectors and firm sizes, it consistently identifies risk-relevant ESG variables with high explanatory power. 
This robustness across different market segments indicates that the framework can be effectively extended to other stock universe. 

\newpage
\bibliographystyle{apalike}
\bibliography{references}

@article{verheyden2016esg,
  title={ESG for all? The impact of ESG screening on return, risk, and diversification},
  author={Verheyden, Tim and Eccles, Robert G and Feiner, Andreas},
  journal={Journal of Applied Corporate Finance},
  volume={28},
  number={2},
  pages={47--55},
  year={2016},
  publisher={Wiley Online Library}
}

@article{gibson2021esg,
  title={ESG rating disagreement and stock returns},
  author={Gibson Brandon, Rajna and Krueger, Philipp and Schmidt, Peter Steffen},
  journal={Financial Analysts Journal},
  volume={77},
  number={4},
  pages={104--127},
  year={2021},
  publisher={Taylor \& Francis}
}

@article{alsaadi2017corporate,
  title={Corporate social responsibility, Shariah-compliance, and earnings quality},
  author={Alsaadi, Abdullah and Ebrahim, M Shahid and Jaafar, Aziz},
  journal={Journal of financial services research},
  volume={51},
  pages={169--194},
  year={2017},
  publisher={Springer}
}

@article{engelhardt2021esg,
  title={ESG ratings and stock performance during the COVID-19 crisis},
  author={Engelhardt, Nils and Ekkenga, Jens and Posch, Peter},
  journal={Sustainability},
  volume={13},
  number={13},
  pages={7133},
  year={2021},
  publisher={MDPI}
}

@article{giese2019foundations,
  title={Foundations of ESG investing: How ESG affects equity valuation, risk, and performance},
  author={Giese, Guido and Lee, Linda-Eling and Melas, Dimitris and Nagy, Zolt{\'a}n and Nishikawa, Laura},
  journal={The Journal of Portfolio Management},
  volume={45},
  number={5},
  pages={69--83},
  year={2019},
  publisher={Portfolio Management Research}
}

@article{fauver2018does,
  title={Does it pay to treat employees well? International evidence on the value of employee-friendly culture},
  author={Fauver, Larry and McDonald, Michael B and Taboada, Alvaro G},
  journal={Journal of Corporate Finance},
  volume={50},
  pages={84--108},
  year={2018},
  publisher={Elsevier}
}

@inproceedings{guthrie2007legitimacy,
  title={Legitimacy theory: A story of reporting social and environmental matters within the Australian food and beverage industry},
  author={Guthrie, James and Cuganesan, Suresh and Ward, Leanne},
  booktitle={Asia Pacific Interdisciplinary Research in Accounting Conference (5th: 2007)},
  pages={1--35},
  year={2007},
  organization={APIRA 2007 Organising Committee}
}

@article{patten1991exposure,
  title={Exposure, legitimacy, and social disclosure},
  author={Patten, Dennis M},
  journal={Journal of Accounting and public policy},
  volume={10},
  number={4},
  pages={297--308},
  year={1991},
  publisher={Elsevier}
}

@article{zhou2021esg,
  title={ESG performance and stock price volatility in public health crisis: evidence from COVID-19 pandemic},
  author={Zhou, Dongyi and Zhou, Rui},
  journal={International Journal of Environmental Research and Public Health},
  volume={19},
  number={1},
  pages={202},
  year={2021},
  publisher={MDPI}
}

@article{gurvich2022carbon,
  title={Carbon risk factor framework},
  author={Gurvich, Alex and Creamer, Germ{\'a}n G},
  journal={The Journal of Portfolio Management},
  volume={48},
  number={10},
  pages={148--164},
  year={2022},
  publisher={Institutional Investor Journals Umbrella}
}

@article{deng2019can,
  title={Can ESG indices improve the enterprises’ stock market performance?—An empirical study from China},
  author={Deng, Xiang and Cheng, Xiang},
  journal={Sustainability},
  volume={11},
  number={17},
  pages={4765},
  year={2019},
  publisher={MDPI}
}

@article{loof2022corporate,
  title={Is Corporate Social Responsibility investing a free lunch? The relationship between ESG, tail risk, and upside potential of stocks before and during the COVID-19 crisis},
  author={L{\"o}{\"o}f, Hans and Sahamkhadam, Maziar and Stephan, Andreas},
  journal={Finance Research Letters},
  volume={46},
  pages={102499},
  year={2022},
  publisher={Elsevier}
}

@article{bonacorsi2024esg,
  title={ESG Factors and Firms' Credit Risk},
  author={Bonacorsi, Laura and Cerasi, Vittoria and Galfrascoli, Paola and Manera, Matteo},
  journal={Journal of Climate Finance},
  volume={6},
  pages={100032},
  year={2024},
  publisher={Elsevier}
}

@article{halbritter2015wages,
  title={The wages of social responsibility—where are they? A critical review of ESG investing},
  author={Halbritter, Gerhard and Dorfleitner, Gregor},
  journal={Review of Financial Economics},
  volume={26},
  pages={25--35},
  year={2015},
  publisher={Elsevier}
}

@article{alareeni2020esg,
  title={ESG impact on performance of US S\&P 500-listed firms},
  author={Alareeni, Bahaaeddin Ahmed and Hamdan, Allam},
  journal={Corporate Governance: The International Journal of Business in Society},
  volume={20},
  number={7},
  pages={1409--1428},
  year={2020},
  publisher={Emerald Publishing Limited}
}

@article{shen2023esg,
  title={ESG in China: A review of practice and research, and future research avenues},
  author={Shen, Hongtao and Lin, Honghui and Han, Wenqi and Wu, Huiying},
  journal={China Journal of Accounting Research},
  pages={100325},
  year={2023},
  publisher={Elsevier}
}

@article{clarkson2008revisiting,
  title={Revisiting the relation between environmental performance and environmental disclosure: An empirical analysis},
  author={Clarkson, Peter M and Li, Yue and Richardson, Gordon D and Vasvari, Florin P},
  journal={Accounting, organizations and society},
  volume={33},
  number={4-5},
  pages={303--327},
  year={2008},
  publisher={Elsevier}
}

@article{cheng2014corporate,
  title={Corporate social responsibility and access to finance},
  author={Cheng, Beiting and Ioannou, Ioannis and Serafeim, George},
  journal={Strategic management journal},
  volume={35},
  number={1},
  pages={1--23},
  year={2014},
  publisher={Wiley Online Library}
}

@article{perez2022does,
  title={Does ESG really matter—and why},
  author={P{\'e}rez, Lucy and Hunt, Vivian and Samandari, Hamid and Nuttall, Robin and Biniek, Krysta},
  journal={McKinsey Quarterly},
  volume={60},
  number={1},
  year={2022}
}

@article{park2024empirical,
  title={An Empirical Analysis on Performance Inconsistency among Environmental, Social and Governance Components of ESG Ratings},
  author={Park, Minjung},
  journal={Asia-Pacific Journal of Business},
  volume={15},
  number={1},
  pages={33--44},
  year={2024},
  publisher={KNU The Institute of Management \& Economy Research}
}

@article{wang2023sparsity,
  title={A sparsity algorithm for finding optimal counterfactual explanations: Application to corporate credit rating},
  author={Wang, Dan and Chen, Zhi and Florescu, Ionu{\c{t}} and Wen, Bingyang},
  journal={Research in International Business and Finance},
  volume={64},
  pages={101869},
  year={2023},
  publisher={Elsevier}
}

@article{Hoerl01021970,
author = {Arthur E. Hoerl and Robert W. Kennard},
title = {Ridge Regression: Biased Estimation for Nonorthogonal Problems},
journal = {Technometrics},
volume = {12},
number = {1},
pages = {55--67},
year = {1970},
publisher = {ASA Website},
doi = {10.1080/00401706.1970.10488634},
}

@article{jain2019can,
  title={Can sustainable investment yield better financial returns: A comparative study of ESG indices and MSCI indices},
  author={Jain, Mansi and Sharma, Gagan Deep and Srivastava, Mrinalini},
  journal={Risks},
  volume={7},
  number={1},
  pages={15},
  year={2019},
  publisher={MDPI}
}

@article{golbayani2020application,
  title={Application of deep neural networks to assess corporate credit rating},
  author={Golbayani, Parisa and Wang, Dan and Florescu, Ionut},
  journal={arXiv preprint arXiv:2003.02334},
  year={2020}
}

@article{eikon2022environmental,
  title={Environmental, social and governance scores from Refinitiv},
  author={Eikon, Refinitiv},
  journal={London: Refinitiv Eikon},
  year={2022}
}

@article{li2021esg,
  title={ESG: Research progress and future prospects},
  author={Li, Ting-Ting and Wang, Kai and Sueyoshi, Toshiyuki and Wang, Derek D},
  journal={Sustainability},
  volume={13},
  number={21},
  pages={11663},
  year={2021},
  publisher={MDPI}
}

@article{gray1995corporate,
  title={Corporate social and environmental reporting: a review of the literature and a longitudinal study of UK disclosure},
  author={Gray, Rob and Kouhy, Reza and Lavers, Simon},
  journal={Accounting, auditing \& accountability journal},
  volume={8},
  number={2},
  pages={47--77},
  year={1995},
  publisher={MCB UP Ltd}
}

@article{benvenuto2023systematic,
  title={A systematic literature review on the determinants of sustainability reporting systems},
  author={Benvenuto, Marco and Aufiero, Chiara and Viola, Carmine},
  journal={Heliyon},
  volume={9},
  number={4},
  year={2023},
  publisher={Elsevier}
}

@article{erhart2022take,
  title={Take it with a pinch of salt—ESG rating of stocks and stock indices},
  author={Erhart, Szil{\'a}rd},
  journal={International Review of Financial Analysis},
  volume={83},
  pages={102308},
  year={2022},
  publisher={Elsevier}
}

@article{sabbaghi2023esg,
  title={ESG and volatility risk: International evidence},
  author={Sabbaghi, Omid},
  journal={Business Ethics, the Environment \& Responsibility},
  volume={32},
  number={2},
  pages={802--818},
  year={2023},
  publisher={Wiley Online Library}
}

@article{livieri2024pricing,
  title={Pricing transition risk with a jump-diffusion credit risk model: evidences from the CDS market},
  author={Livieri, Giulia and Radi, Davide and Smaniotto, Elia},
  journal={Review of Corporate Finance},
  volume={4},
  number={1},
  pages={177--201},
  year={2024},
  publisher={Emerald Publishing Limited}
}

@article{zhang2025impact,
  title={The Impact of Controversial Events on Corporate Resilience: The Chain-Mediating Role of ESG and Value-at-Risk},
  author={Zhang, Jie and Wang, Derek D},
  journal={Sustainability},
  volume={17},
  number={24},
  pages={11032},
  year={2025},
  publisher={MDPI}
}

@article{bolton2023global,
  title={Global pricing of carbon-transition risk},
  author={Bolton, Patrick and Kacperczyk, Marcin},
  journal={The Journal of Finance},
  volume={78},
  number={6},
  pages={3677--3754},
  year={2023},
  publisher={Wiley Online Library}
}

@article{pastor2022dissecting,
  title={Dissecting green returns},
  author={P{\'a}stor, L'ubo{\v{s}} and Stambaugh, Robert F and Taylor, Lucian A},
  journal={Journal of financial economics},
  volume={146},
  number={2},
  pages={403--424},
  year={2022},
  publisher={Elsevier}
}

@article{wei2006did,
  title={Why did individual stocks become more volatile?},
  author={Wei, Steven X and Zhang, Chu},
  journal={The Journal of Business},
  volume={79},
  number={1},
  pages={259--292},
  year={2006},
  publisher={JSTOR}
}

@article{al2021firm,
  title={Firm-specific, macroeconomic factors and stock price risk for Jordanian banks},
  author={Al Salamat, Wasfi and Momani, Mohammad QM and Batayneh, Khaled},
  journal={Banks and Bank Systems},
  volume={16},
  number={3},
  pages={166},
  year={2021},
  publisher={Business Perspectives Ltd.}
}

@article{pae2018idiosyncratic,
  title={Idiosyncratic volatility and cash flow volatility: New evidence from S\&P 500},
  author={Pae, Yuntaek and Bae, Sung C and Lee, Namhoon},
  journal={International Review of Financial Analysis},
  volume={56},
  pages={127--135},
  year={2018},
  publisher={Elsevier}
}

@article{konar2001does,
  title={Does the market value environmental performance?},
  author={Konar, Shameek and Cohen, Mark A},
  journal={Review of economics and statistics},
  volume={83},
  number={2},
  pages={281--289},
  year={2001},
  publisher={MIT Press 238 Main St., Suite 500, Cambridge, MA 02142-1046, USA journals~…}
}

@article{matsumura2014firm,
  title={Firm-value effects of carbon emissions and carbon disclosures},
  author={Matsumura, Ella Mae and Prakash, Rachna and Vera-Mu{\~n}oz, Sandra C},
  journal={The accounting review},
  volume={89},
  number={2},
  pages={695--724},
  year={2014},
  publisher={American Accounting Association}
}


\newpage

\appendix

\section{Data Processing}
\label{appendix:dataprocessing}
First, we exclude demographic variables, such as CEO names, which fall outside the scope of this study. We then handle missing boolean variables and numeric variables. 
Following \citep{clarkson2008revisiting, cheng2014corporate}, we assign 0 to missing values for all boolean variables. \cite{clarkson2008revisiting} examine boolean variables that indicate whether companies issue policy, regulation, or measures related to ESG. These are coded ``1'' if present, ``0'' otherwise in our dataset. The authors infer that if companies issue ESG related policies, they are likely to report their efforts to the public, thus the missing values can be reasonably assessed as no such regulations are issued. \cite{cheng2014corporate} examine boolean variables that indicate whether companies are involved in events or practices considered adverse from an ESG perspective (e.g., weapons manufacturing, alcohol, etc.). They believe that if companies are involved in adverse practices, the public media is likely to report it, so the missing values can be considered as they are not involved. We carefully examined each variable description in the LSEG dataset, and we found that all boolean variables fit one of these two cases. 

For numeric variables, we first consider ``controversy variables'' that quantify regulatory violations. Similar to our treatment of adverse event boolean variables, we assume missing controversy data implies no significant incidents occurred, and thus impute these missing values as ``0''. For the remaining numeric variables, we retain only those variables with at least 80\% data availability within each sector across the entire analysis period. This step is performed separately with respect to each sector as we find data availability across industries is different. 

Finally, any observation (i.e., company-year data point) still containing missing values after these steps is removed from the dataset. Taking the Energy sector as an example, the initial dataset had 617 variables and 695 company/year observations. After applying the data processing steps, the final dataset for the Energy sector contains 255 variables and 422 observations.

\section{Applying HVS on Returns}
\label{appendix:applyinghvsonreturns}
When applying our algorithm to model returns, we find that the resulting \%dev values are significantly lower compared to those obtained when modeling risk. The HVS algorithm identifies significantly fewer ESG variables as relevant for explaining returns than for risk. For example, in the Energy sector, it selects only six variables for returns compared to 31 for risk. These six variables are: Global Compact, Human Rights Processes in Policy Forced Labor, Sdg17PartnershipsToAchieveTheGoal, Veto Power or Golden share, Sdg13ClimateAction, Human Rights Process in Policy Freedom of Association.

\begin{figure}[htp]
    \makebox[\textwidth][c]{
        \begin{minipage}{1\textwidth}  
            \centering
            \caption{This bar chart illustrates the \%dev of using selected raw variables within each category to regress returns.}
            \label{returnbarchartforstep1}
            \includegraphics[width=\linewidth]{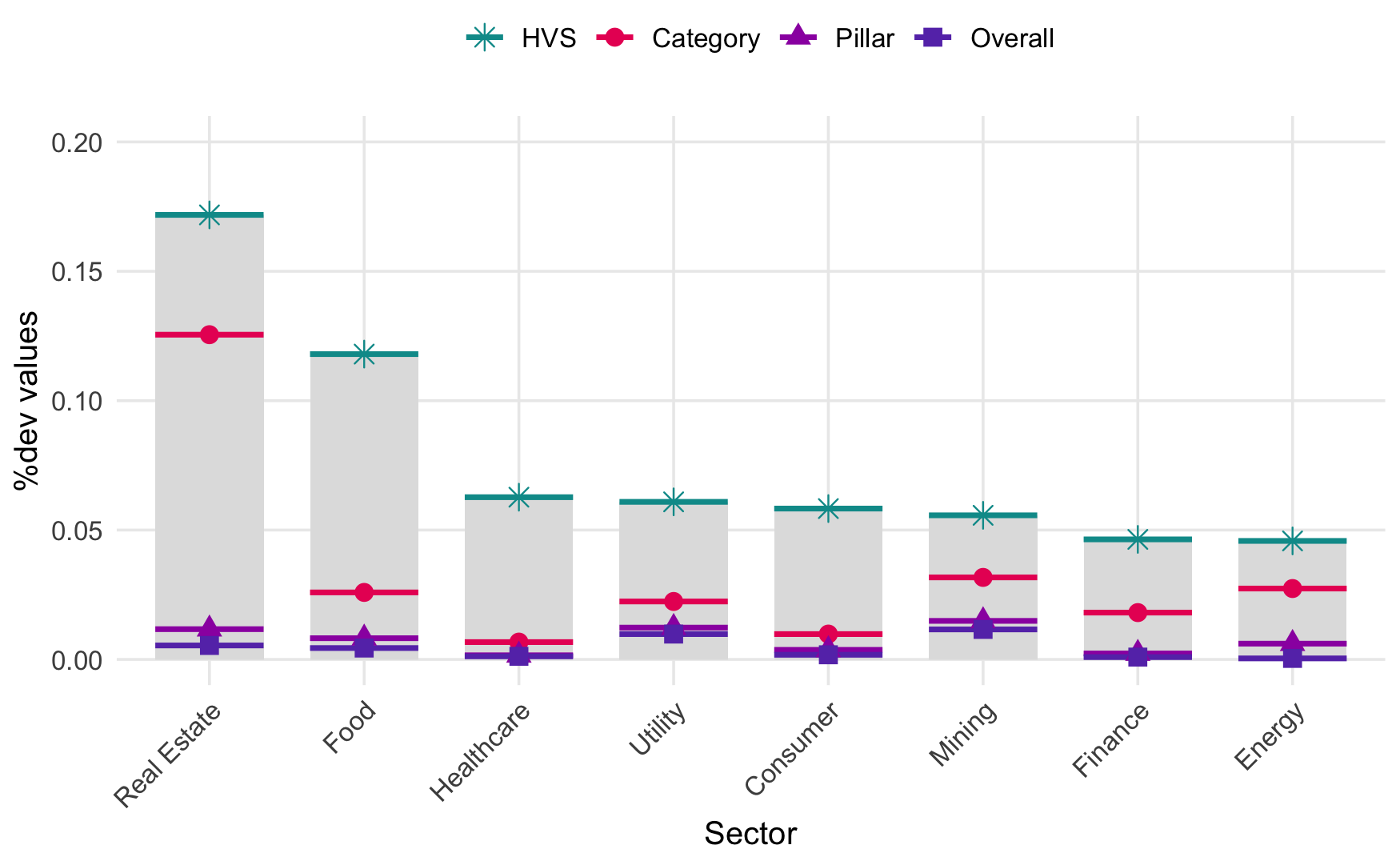}
        \end{minipage}
    }
\end{figure}

These findings illustrate that ESG factors are more related to risk rather than return, which confirms findings from \cite{perez2022does, engelhardt2021esg, deng2019can, fauver2018does}.

\section{Validation For Other Sectors}
\label{appendix:validationforothersectors}

\begin{table}[htbp]
  \centering
  \caption{Comparison of HVS Performance with Benchmark Methods for Each Sector}
  \footnotesize
    \begin{tabular}{lrrrr}
    \textbf{Food} &       &       &       &  \\
    \midrule
    \textbf{Benchmark} & \multicolumn{1}{l}{\textbf{No.Selected Variables}} & \multicolumn{1}{l}{\textbf{\%dev}} & \multicolumn{1}{l}{\textbf{AIC}} & \multicolumn{1}{l}{\textbf{BIC}} \\
    \midrule
    PCA   & 53    & 0.53  & 217.01 & 435.29 \\
    PCA   & 23    & 0.41 & 240.59 & 339.81 \\
    Stepwise & 62    & 0.56  & 132.81 & 382.84 \\
    Lasso & 52    & 0.51  & 224.70 & 437.04 \\
    \midrule
    HVS   & 23    & 0.55  & 136.10 & 233.35 \\
    \midrule
          &       &       &       &  \\
    \textbf{Mining} &       &       &       &  \\
    \midrule
    \textbf{Model} & \multicolumn{1}{l}{\textbf{No.Selected Variables}} & \multicolumn{1}{l}{\textbf{\%dev}} & \multicolumn{1}{l}{\textbf{AIC}} & \multicolumn{1}{l}{\textbf{BIC}} \\
    \midrule
    PCA   & 33    & 0.43  & 153.18 & 270.32 \\
    PCA   & 21    & 0.28 & 177.69 & 254.68 \\
    Stepwise & 32    & 0.22  & 180.26 & 290.71 \\
    Lasso & 39    & 0.50  & 134.90 & 274.14 \\
    HVS   & 21    & 0.52  & 89.52 & 165.15 \\
    \midrule
          &       &       &       &  \\
    \textbf{Healthcare} &       &       &       &  \\
    \midrule
    \textbf{Benchmark} & \multicolumn{1}{l}{\textbf{No.Selected Variables}} & \multicolumn{1}{l}{\textbf{\%dev}} & \multicolumn{1}{l}{\textbf{AIC}} & \multicolumn{1}{l}{\textbf{BIC}} \\
    \midrule
    PCA   & 66    & 0.41  & 620.36 & 930.22 \\
    PCA   & 40    & 0.37  & 621.00 & 812.39 \\
    Stepwise & 67    & 0.55  & 351.60 & 661.47 \\
    Lasso & 39    & 0.41  & 559.86 & 744.13 \\
    \midrule
    HVS   & 40    & 0.51  & 440.83 & 629.65 \\
    \midrule
          &       &       &       &  \\
    \textbf{Industry} &       &       &       &  \\
    \midrule
    \textbf{Benchmark} & \multicolumn{1}{l}{\textbf{No.Selected Variables}} & \multicolumn{1}{l}{\textbf{\%dev}} & \multicolumn{1}{l}{\textbf{AIC}} & \multicolumn{1}{l}{\textbf{BIC}} \\
    \midrule
    PCA   & 75    & 0.44  & 541.80 & 905.44 \\
    PCA   & 38    & 0.30  & 649.47 & 838.38 \\
    Stepwise & 75    & 0.52  & 321.83 & 680.75 \\
    Lasso & 65    & 0.43  & 520.81 & 834.51 \\
    \midrule
    HVS   & 38    & 0.44  & 460.58 & 646.76 \\
    \bottomrule
    \end{tabular}
  \label{tab:Validation For Other Sectors 1}
\end{table}

\begin{table}[htbp]
  \centering
  \caption{Comparison of HVS Performance with Benchmark Methods for Each Sector}
  \footnotesize
    \begin{tabular}{lrrrr}
    \textbf{Real Estate} &       &       &       &  \\
    \midrule
    \textbf{Benchmark} & \multicolumn{1}{l}{\textbf{No.Selected Variables}} & \multicolumn{1}{l}{\textbf{\%dev}} & \multicolumn{1}{l}{\textbf{AIC}} & \multicolumn{1}{l}{\textbf{BIC}} \\
    \midrule
    PCA   & 30    & 0.25  & 291.31 & 395.05 \\
    PCA   & 11    & 0.19  & 268.36 & 310.51 \\
    Stepwise & 34    & 0.58  & 152.56 & 266.02 \\
    Lasso & 2     & 0.07  & 271.95 & 283.68 \\
    \midrule
    HVS   & 11    & 0.42  & 200.11 & 241.01 \\
    \midrule
          &       &       &       &  \\
    \textbf{Utility} &       &       &       &  \\
    \midrule
    \textbf{Benchmark} & \multicolumn{1}{l}{\textbf{No.Selected Variables}} & \multicolumn{1}{l}{\textbf{\%dev}} & \multicolumn{1}{l}{\textbf{AIC}} & \multicolumn{1}{l}{\textbf{BIC}} \\
    \midrule
    PCA   & 56    & 0.37  & 400.61 & 641.1 \\
    PCA   & 30    & 0.27 & 411.57 & 544.26 \\
    Stepwise & 56    & 0.42  & 300.24 & 536.58 \\
    Lasso & 17    & 0.24  & 405.76 & 482.40 \\
    \midrule
    HVS   & 30    & 0.42  & 304.28 & 434.81 \\
    \midrule
          &       &       &       &  \\
    \textbf{Consumer} &       &       &       &  \\
    \midrule
    \textbf{Model} & \multicolumn{1}{l}{\textbf{No.Selected Variables}} & \multicolumn{1}{l}{\textbf{\%dev}} & \multicolumn{1}{l}{\textbf{AIC}} & \multicolumn{1}{l}{\textbf{BIC}} \\
    \midrule
    PCA   & 58    & 0.35  & 414.85 & 662.46 \\
    PCA   & 27    & 0.22 & 436.24 & 555.92 \\
    Stepwise & 55    & 0.49  & 240.67 & 471.77 \\
    Lasso & 37    & 0.35  & 369.37 & 528.19 \\
    \midrule
    HVS   & 27    & 0.42  & 301.27 & 418.82 \\
    \midrule
          &       &       &       &  \\
    \textbf{Finance} &       &       &       &  \\
    \midrule
    \textbf{Benchmark} & \multicolumn{1}{l}{\textbf{No.Selected Variables}} & \multicolumn{1}{l}{\textbf{\%dev}} & \multicolumn{1}{l}{\textbf{AIC}} & \multicolumn{1}{l}{\textbf{BIC}} \\
    \midrule
    PCA   & 51    & 0.18  & 901.22 & 1160.8 \\
    PCA   & 20    & 0.06  & 972.38 & 1080.1 \\
    Stepwise & 50    & 0.28  & 717.19 & 957.18 \\
    Lasso & 4     & 0.11  & 881.80 & 908.29 \\
    \midrule
    HVS   & 20    & 0.23  & 769.72 & 874.57 \\
    \bottomrule
    \end{tabular}
  \label{tab:Validation For Other Sectors 1}
\end{table}

\newpage
\section{Definition of Risk Variables}
\label{appendix:definitionofriskvariables}


\footnotesize
\begin{longtable}{p{4.5cm}p{9cm}}
\caption{Definition of Energy Risk Variables} \label{tab:definitionofenergyriskvariables} \\
\toprule
\textbf{Variable Name} & \textbf{Definition} \\
\midrule
\endfirsthead

\multicolumn{2}{c}%
{\tablename\ \thetable\ -- \textit{Continued from previous page}} \\
\toprule
\textbf{Variable Name} & \textbf{Definition} \\
\midrule
\endhead

\midrule
\multicolumn{2}{r}{\textit{Continued on next page}} \\
\endfoot

\bottomrule
\endlastfoot

PolicyCyberSecurity & 
Does the company have policies on protecting personal information, ensuring secure data handling practices, and mitigating information security risks? \\
\addlinespace

Total Senior Executives Compensation & 
The total compensation paid to all senior executives (if total aggregate is reported by the company). \\
\addlinespace

Board Size & 
The total number of board members at the end of the fiscal year. \\
\addlinespace

Compensation Policy Elements/Policy Executive Retention & 
Does the company have a compensation policy to attract and retain executives? \\
\addlinespace

Management Training) & 
Does the company claim to provide regular staff and business management training for its managers? \\
\addlinespace

Staff Transport Impact Reduction Initiatives & 
Does the company report on initiatives to reduce the environmental impact of transportation used for its staff? \\
\addlinespace

Armaments 5\% Revenues & 
Are revenues generated from armaments larger than 5\% of the total net revenues? \\
\addlinespace

Estimated CO2 Equivalents Emission Total & 
The estimated total CO2 and CO2 equivalents emission in tonnes. \\
\addlinespace

External Consultants & 
Does the board or board committees have the authority to hire external advisers or consultants without management's approval? \\
\addlinespace

Limited Shareholder Rights to Call Meetings & 
Has the company limited the rights of shareholders to call special meetings? \\
\addlinespace

Balanced Board Structure Policy Elements/Policy Board Size & 
Does the company have a policy regarding the size of its board? \\
\addlinespace

CEO Compensation Link to Total Shareholder Return & 
Is the CEO's compensation linked to total shareholder return (TSR)? \\
\addlinespace

Written Consent Requirements & 
Does the company permit actions to be taken without meeting by written consent? \\
\addlinespace

e-Waste Reduction Initiatives & 
Does the company report on initiatives to recycle, reduce, reuse, substitute, treat or phase out e-waste? \\
\addlinespace

Environmental Products & 
Does the company report on at least one product line or service that is designed to have positive effect on the environment or which is environmentally labeled and marketed? \\
\addlinespace

Critical Countries Controversies & 
Number of controversies published in the media linked to activities in critical, undemocratic countries that do not respect fundamental human rights principles. \\
\addlinespace

OilAndGasProducerFlag & 
Indicator used to identify companies involved in oil and gas production \\
\addlinespace

CSR Sustainability External Audit & 
Does the company monitor its integrated strategy through belonging to a specific sustainability index? AND Does the company monitor its integrated strategy through conducting external audits on its reporting? \\
\addlinespace

Renewable Energy Use & 
Does the company make use of renewable Energy? \\
\addlinespace

Shareholder Rights Policy Elements/Policy Shareholder Engagement & 
Does the company have a policy to facilitate shareholder engagement, resolutions or proposals? \\
\addlinespace

Management Departures & 
Has an important executive management team member or a key team member announced a voluntary departure (other than for retirement) or has been ousted? \\
\addlinespace

Corporate Responsibility Awards & 
Has the company received an award for its social, ethical, community, or environmental activities or performance? \\
\addlinespace

Fundamental Human Rights ILO or UN & 
Does the company have a policy to guarantee the freedom of association universally applied independent of local laws? AND Does the company have a policy for the exclusion of child, forced or compulsory labour? \\
\addlinespace

Whistleblower Protection & 
Does the company describe the implementation of its community policy through a public commitment from a senior management or board member? AND Does the company describe the implementation of its community policy through the processes in place? \\
\addlinespace

Unlimited Authorized Capital or Blank Check & 
Does the company have unlimited authorized capital or a blank check? \\
\addlinespace

Elimination of Cumulative Voting Rights & 
Has the company reduced or eliminated cumulative voting in regard to the election of board members? \\
\addlinespace

OECD Guidelines for Multinational Enterprises & 
Does the company have a policy to improve customer satisfaction or to strive to be a fair competitor? \\
\addlinespace

Confidential Voting Policy & 
Does the company have a confidential voting policy (i.e., management cannot view the results of shareholder votes)? \\
\addlinespace

Diversity and Opportunity Processes/Policy Diversity and Opportunity & 
Does the company have a policy to drive diversity and equal opportunity? \\
\addlinespace

Poison Pill & 
Does the company have a poison pill (shareholder rights plan, macaroni defense, etc.)? \\
\addlinespace

Agrochemical Products & 
Does the company produce or distribute agrochemicals like pesticides, fungicides or herbicides? \\

\end{longtable}

\section{Residual Analysis}
\begin{figure}
    \centering
    \includegraphics[width=1\linewidth]{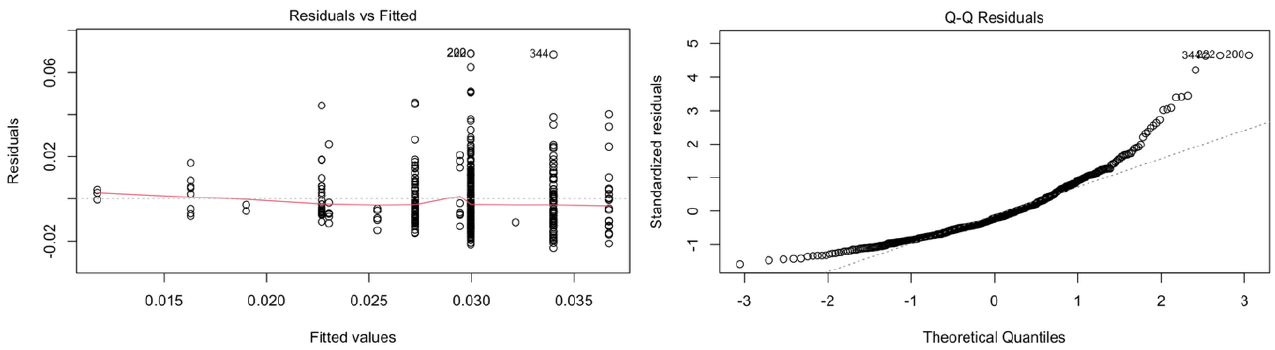}
    \caption{Residual analysis}
    \label{fig:residual_analysis}
\end{figure}

\section{Sector by size analysis}
\begin{figure}
    \centering
    \includegraphics[width=1\linewidth]{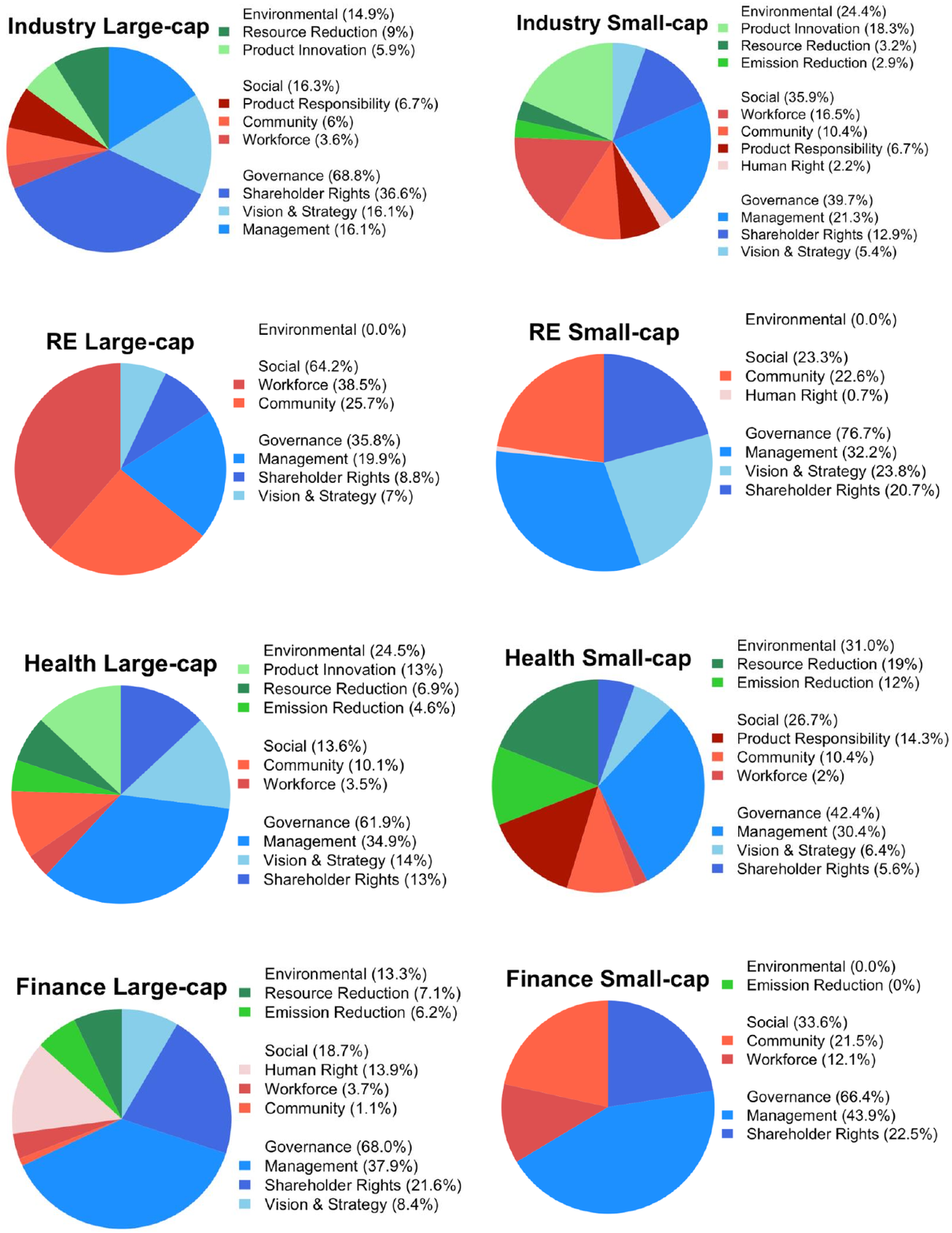}
    \caption{ESG variables within each sector for large and small cap companies.}
    \label{fig:pattern2}
\end{figure}

\end{document}